**From Halos to Galaxies XII.**

**A Unified Explanation for JWST Little Red Dots and High-Redshift Low-Mass Disk-like Galaxies: Prolate Galaxies Viewed End-on vs Side-on**

Yingjie Peng[1,2]
1. Department of Astronomy, School of Physics, Peking University, Beijing, China.
2. Kavli Institute for Astronomy and Astrophysics (KIAA), Peking University, Beijing, China.
yjpeng@pku.edu.cn

**ABSTRACT**
Recent JWST surveys have revealed two puzzling high-redshift phenomena: (1) an unexpectedly large abundance of flattened, disk-like galaxies at z > 3, and (2) a rare population of compact, extremely red sources at z ~ 4–9 ("Little Red Dots," LRDs) that often show V-shaped SEDs and very broad Balmer lines. These findings lack a consensus interpretation and have motivated models ranging from dusty starbursts to obscured AGN and more exotic scenarios. We propose that both phenomena are linked by a simple geometric consequence of a third clue: mounting evidence from structural modeling and axis-ratio statistics indicates that many low-mass galaxies at z > 3 are intrinsically prolate (cigar-like), not oblate rotation-supported disks. In this picture, a substantial fraction of the flattened, disk-like morphologies reported at z > 3 arise from side-on and intermediate-angle projections of prolate systems, while the rare near end-on views appear extremely compact and high-surface-brightness, and are preferentially reddened by the maximal line-of-sight column, naturally matching key elements of LRD selection. The expected fraction of near end-on systems, $P_{end} = 1 - \cos(\theta_{max})$, is ~1–3% for $\theta_{max}$ ~ 10–15°, consistent with LRD demographics in wide JWST fields. This orientation-based framework does not exclude AGN or starburst activity; rather, it explains LRD rarity as an orientation effect and provides a natural route to the large columns of gas/dust and scattering depths inferred in recent dense-gas and electron-scattering interpretations of LRD spectra, without fine-tuned new physics. The model makes falsifiable predictions to validate or rule out this geometric interpretation.

## 1. Introduction

JWST has dramatically improved our view of galaxy morphology and nuclear activity at early cosmic times. Recent JWST observations have revealed two puzzling phenomena in the high-redshift universe that challenge our understanding of galaxy formation.

Abundance of disk-like galaxies at early times: Contrary to expectations that early galaxies would be irregular and dynamically hot, JWST found a surprisingly large number of flattened, disk-like systems at z > 3 (e.g. Ferreira et al. 2022). These galaxies appear to have thin, extended shapes reminiscent of disks, even in the first 1–2 billion years of cosmic history. This was unexpected because traditional models (tidal torque theory, etc.) suggest early galaxies form via chaotic mergers and only settle into well-ordered disks later in cosmic time.

Little Red Dots (LRDs): JWST deep fields also uncovered a rare population of compact, extremely red objects at z~4–9, dubbed Little Red Dots (e.g., Kokorev et al. 2024; Kocevski et al. 2025; Akins et al. 2025; Inayoshi & Ho 2025 for a review). LRDs have very small observed sizes (effective radii of order tens to a few hundred parsecs), unusually red colors (often with a prominent Balmer break in their spectrum), and broad emission lines indicating large velocity dispersions (FWHM of several thousand km/s) (e.g. Kokorev et al. 2024; Kocevski et al. 2025; Rusakov et al. 2026). They are rare, occupying a percent-level regime relative to the general star-forming population at similar redshifts and brightness, yet generally far more numerous than UV-bright quasars at similar epochs. LRDs thus occupy an intermediate number density regime, making up on the order of only a few percent of galaxies in the z~4–9 epoch (Kokorev et al. 2024; Akins et al. 2025). Their rarity and extreme properties have been deemed a major challenge to existing galaxy evolution theories.

These two findings: abundant high-z disk-like galaxies and the rare LRDs, lack a widely accepted explanation and are often seen as challenges to standard expectations and implementations of galaxy formation models within ΛCDM framework. In the case of LRDs, numerous exotic models have been proposed: e.g. self-interacting dark matter halos, primordial black hole seeding, new black hole accretion physics, modifications to dark energy or cosmic expansion, etc. These models attempt to address one or another facet of LRDs, but none fully explain the combination of observed properties: the extreme compactness, their rarity, high velocity dispersions, broad lines, a V-shaped SED in which the rest-UV remains detectable while the rest-optical becomes extremely red, together with tight far-IR/submm emission constraints for many samples (Kokorev et al. 2024; Casey et al. 2025). For example, interpreting LRDs as dust-obscured starbursts or moderate-luminosity AGN runs into problems: heavy dust extinction would produce far-infrared thermal emission that isn't observed (Casey et al. 2025), and AGN models struggle with the fact that LRDs are X-ray faint compared to low-redshift type-I AGN scaling relations (Yue et al. 2024) despite their broad lines, suggesting any central black holes must be either extremely obscured or accreting in an unusual mode.

In this work, we propose that these two apparently separate JWST discoveries: abundant high-z disk-like galaxies and rare LRDs, may be linked by a simple and testable geometric consequence of the intrinsic shapes of early low-mass galaxies.

## 2. Evidence for Ubiquitous Prolate Low-mass Galaxies at High Redshift

A key observational development enabled by JWST suggests a unifying clue: the shapes and kinematics of many high-z galaxies indicate they are neither classic disks nor irregular spheroids, but elongated (prolate) structures. Recent studies have found that low-mass galaxies at z>3 often have very high ellipticities (axis ratios $b/a \sim 0.2$–$0.4$ and $c/a \sim 0.2$ are common, e.g. Pandya et al. 2024), and their overall axis-ratio distribution strongly deviates from that of local galaxies. In fact, the observed shape distribution at z~3–8 matches what is expected for randomly oriented prolate ellipsoids (cigar-like shapes), and is inconsistent with an oblate (disk-dominated) population (Wang et al. 2024).

Pandya et al. (2024) use JWST/CEERS imaging to infer the 3D geometry of high-redshift galaxies and found that the fraction of dwarf galaxies with prolate (elongated) shapes rises from ~25% at z~0.5 to 50–80% at z=3–8. In other words, most low-mass galaxies in the early universe might have been elongated rather than disk-shaped. This surprising result follows strong statistical support: Wang et al. (2024) show that the ellipticity distribution of JWST galaxies with stellar mass $M^* \sim 10^9$-$10^{9.5}$ $M_\odot$ at z=3–8 is well fit by a prolate model, whereas local galaxies of similar mass follow an oblate (disky) distribution. This indicates a shape transition from prolate at high-z to oblate at low-z, likely reflecting a change in orbital anisotropy (from radial orbits at high-z to more circular orbits at low-z as galaxies settle). Furthermore, these high-z elongated galaxies are not necessarily rotation-supported "cold disks" at all, kinematic analysis suggests they are likely dynamically warm/hot systems dominated by random motions (with flattening due to anisotropy rather than rotation). In summary, JWST data point to a paradigm where many early galaxies start out as triaxial or prolate structures aligned with cosmic web filaments, challenging the assumption that all galaxies begin as classic rotating disks.

An important concern is whether selection effects could artificially boost the observed number of elongated galaxies at high redshift. For example, edge-on oblate disks and elongated shapes have higher projected surface brightness than face-on disks at fixed total flux, making them easier to detect. However, Pandya et al. (2024) explicitly tested completeness and concluded that for the brightness regime relevant to their primary sample (e.g., galaxies brighter than about 26.5 AB mag in F277W), the JWST imaging is nearly complete even for large, face-on disks. In other words, if a significant population of face-on, round disks existed at high-z, the survey should have detected them. The fact that most of the detected low-mass galaxies are highly elongated is thus a genuine feature, not just a selection artifact. Thus, while projection can influence detectability, current evidence suggests that the strong preference for high ellipticities among low-mass high-z galaxies is not solely an artifact of selection. Moreover, the maximum implied weak-lensing shear also cannot explain the elongated shapes of early galaxies (Pandya et al. 2025).

Theoretically, some simulations confirm that majority of the low-mass galaxies at high redshifts are not discs or spheroids but rather galaxies with elongated morphologies, residing in dark matter (DM) haloes with strongly elongated prolate shape, a common feature of high-redshift DM haloes in the ΛCDM cosmology (Allgood et al. 2006; Ceverino et al. 2015; Tomassetti et al. 2016; Lapiner et al. 2023). Recent hydrodynamical simulations also connects the observed prolate galaxies at high redshift to smooth filamentary assembly, in which coherent inflows along the cosmic web naturally produce intrinsically prolate, elongated shape (Pozo et al. 2026).

Taken together, these results motivate an interpretation in which many low-mass galaxies at z > 3 are intrinsically prolate and dynamically warm/hot rather than classic rotation-supported oblate disks.

## 3. Prolate Galaxies as a Unified Explanation

If low-mass prolate galaxies are ubiquitous at z > 3, then two observational consequences become difficult to avoid: the same intrinsic population will present very different morphologies and columns depending on viewing angle, naturally producing both (i) abundant disk-like projections and (ii) rare compact red end-on projections.

Here we focus on the low-mass regime where prolate-shape evidence is strongest (Pandya et al. 2024; Wang et al. 2024); the model does not rule out that many (or even the vast majority of) higher-mass systems at similar redshifts are genuine rotation-supported disks. As shown in Figure 1, for prolate galaxies, thinking of a stretched rugby ball or cigar shape, the major axis is much longer than the other two axes. Depending on how this elongated galaxy is oriented to our line of sight, it can look very different.

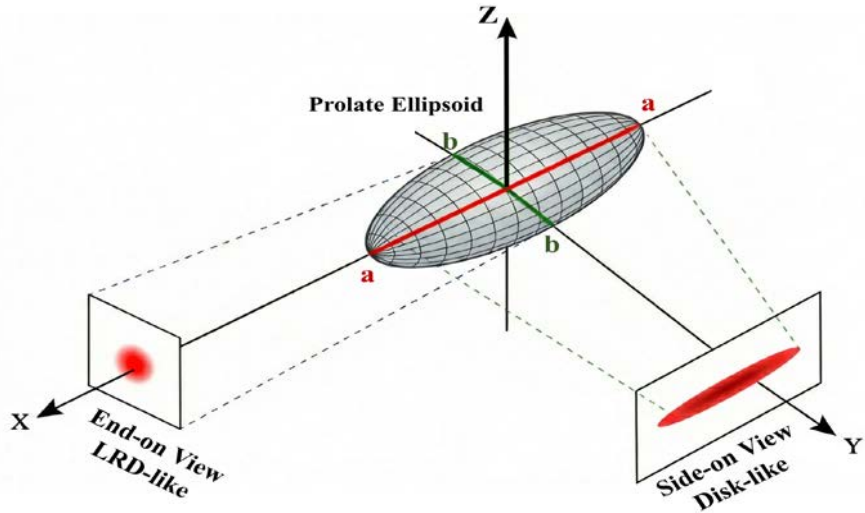

**Figure 1.** Schematic of a unified explanation for JWST Little Red Dots (LRDs) and high-redshift low-mass disk-like galaxies: prolate galaxies viewed end-on vs side-on. A prolate galaxy (a > b ~ c) is shown with its axis indicated. Two representative lines of sight are illustrated. In the end-on configuration (view aligned with the major axis), the galaxy projects to a compact, nearly point-like source with high apparent surface brightness (LRD-like appearance), and the line of sight traverses the maximum dust/gas column, enhancing effective attenuation/absorption and scattering signatures. In the side-on configuration (view perpendicular to the major axis), the same galaxy projects to a flattened, elongated morphology that can be classified as disk-like in imaging (morphological sense), with a shorter typical line-of-sight column.

If we view the galaxy edge-on (side-on), i.e. looking perpendicular to its long axis, we see an extended, flattened shape on the sky. The galaxy's light is spread out along its major axis, resembling a thin disk in projection. This scenario corresponds to the numerous disk-like galaxies JWST has found (e.g., Ferreira et al. 2022). A prolate galaxy seen side-on would have a high apparent ellipticity (elongated profile) and can even show an exponential-like light profile, mimicking a disk (Pandya et al. 2024).

If we view the galaxy end-on (along the long axis), looking down the barrel of the cigar, we see only the small cross-sectional area of the galaxy. The object will appear extremely compact, even point-like, since we are looking at the narrow end. Most of the galaxy's stars and gas are hidden behind one another along the line of sight. Moreover, the line of sight pierces through the maximum depth of the galaxy's interstellar medium. This means the light we receive has traveled through a long path of dust and gas, causing significant extinction and reddening. This scenario, a prolate galaxy viewed end-on would appear as a "little red dot": very compact with high surface brightness, very red, and, as we detail below, exhibiting large apparent velocity dispersion and other observed LRDs features due to the geometry.

Thus, a population of intrinsically prolate galaxies ***inevitably*** produces many flattened "disk-like" objects (side-on and intermediate angles) and a small subset of extremely compact sources (near end-on). This single explanation ties together two seemingly unrelated phenomena without invoking new exotic physics. It shifts the perspective from "JWST is seeing surprisingly many disks and also some weird red compact objects" to "JWST is primarily seeing elongated galaxies, most at a tilt, a few end-on." Many puzzling aspects of LRDs become straightforward consequences of geometry and orientation in this scenario.

It is worth noting that this end-on prolate model does not necessarily exclude the presence of AGN or starburst activity in LRDs. An elongated galaxy could very well host a growing central black hole and/or intense star-forming clumps/clusters, the difference is that due to the unique viewing angle, the impact of dust/gas and projection on observed properties is maximal. As we explain below, the model can accommodate a moderate-luminosity AGN at the center of an LRD (contributing to broad lines), without requiring the AGN to have extreme intrinsic luminosity or unusual accretion physics.

## 4. How the End-on Prolate Model Explains LRD Properties

### 4.1 The abundance of LRDs:

As revealed by observations, if ~50–70% of low-mass high-z galaxies are prolate, then statistically a few percent will have their long axis fortuitously aligned with our line of sight. Geometrically, the probability of viewing a randomly oriented object within an angle $\theta_{max}$ of the end-on direction is $P_{end} = 1 - \cos(\theta_{max})$. If we define "end-on" as within say $\theta_{max} \sim 10$ degrees of the exact long-axis alignment, then $P_{end} \sim 1.5\%$. For $\theta_{max} = 15$ degrees, $P_{end} \sim 3.4\%$. Thus, on the order of a few percent of prolate galaxies would be seen as LRD-type objects, consistent with the observed fraction of LRDs (1–5% of galaxies in the relevant redshift and luminosity range, Kokorev et al. 2024; Akins et al. 2025). This naturally explains the rarity of LRDs: they are simply the tail of an orientation distribution. By contrast, a much larger fraction of prolate galaxies will be viewed at intermediate angles or side-on, making them appear as elongated "disky" galaxies (which indeed seem to be common at high-z). The model quantitatively reproduces the LRD abundance without needing fine-tuned new physics: it's an orientation effect on an otherwise fairly common population.

**4.2 Extreme Compactness and Surface Brightness**

An LRD has an observed effective radius of order tens to a few hundred parsecs; often unresolved in JWST images. In our model, this is because we are looking almost directly down the long axis of a prolate galaxy. The entire stellar distribution, which might extend over several kpc in length, is foreshortened into a very small angular size. The projected area is minimal. Thus, even if the galaxy is physically extended, its apparent size is tiny when oriented end-on. This naturally reproduces the compact appearance of LRDs without requiring an actual tiny, disky galaxy.

Another consequence is that the surface brightness of the object remains high. All the light of the galaxy is concentrated in a small image area, meaning high flux per unit area. This "core domination" can make the central region so bright that it drowns out any extended fuzz around it. JWST's point spread function (PSF) will scatter some light into wings around the point-like core, possibly obscuring a faint halo if present. Essentially, an end-on prolate galaxy behaves like a bright central source with a faint extended component that's hard to detect against the glare and background noise. This dynamic range problem is analogous to looking at a distant flashlight: you see the bright point of light, but not the person holding it because they're much fainter compared to the light. In LRDs, any normal outer stellar component would be outshone by the blazing central concentration and rendered invisible in current images. This could explain why LRDs often look like unresolved red points, and the underlying host galaxy light is hard to discern, not necessarily absent.

**4.3 Red Colors and V-shaped SED**

LRDs exhibit extremely red observed near-infrared colors, typically showing a strong flux enhancement toward the F444W band relative to shorter-wavelength filters. Their broadband SEDs often show a pronounced curvature that resembles an apparent Balmer/4000 Å break-like feature, producing a characteristic V-shaped appearance: the rest-UV can remain detectable (and often blue), while the rest-optical rises steeply to very red colors. In standard interpretations, such features could, in principle, be explained by an evolved stellar population or heavy dust attenuation, but both scenarios face challenges. An old stellar population would require very early and rapid star formation at very high redshift, while a single, uniform heavy dust screen must simultaneously explain the very red optical continuum and the fact that many LRDs still show a bright, only weakly attenuated UV component, in addition to the lack of strong far-infrared detections.

Our model offers a geometric explanation. Along the long axis, the line of sight traverses the maximum column of the galaxy's interstellar medium. We are effectively peering through a large fraction of the galaxy's dust and/or dense gas structure. This long path length means that even a moderate volume density can integrate to a high optical depth in projection. Crucially, however, an end-on prolate view does not require the UV to be globally extinguished. High-redshift low-mass galaxies are typically star-forming and clumpy, and the near-side star-forming regions can contribute UV light that reaches the observer through relatively low-column sightlines, while emission originating deeper in the system (or on the far side) is viewed through much larger columns and is strongly reddened or absorbed. This naturally produces strong differential obscuration: the UV can remain visible and blue, while the UV-to-optical ratio is suppressed and the rest-optical continuum appears extremely red. Interestingly, this "two-component / spatially separated" picture is not merely conceptual. A strongly lensed, spatially resolved LRD at z = 6.027 reveals a compact red dot and a spatially offset blue UV component separated by about 300 pc in the source plane; the authors argue that such components would blend into a single compact source with canonical LRD colors in unlensed observations, and that the V-shaped SED can arise from the superposition of two physically distinct components rather than from a single uniformly obscured compact emitter (Baggen et al. 2025).

Crucially, because the obscuring material is distributed over a large volume (the length of the galaxy) rather than concentrated in a compact torus, it can block light without itself reaching extremely high temperatures. In a classical AGN scenario, a compact, optically thick dusty structure near the nucleus would be expected to be heated and emit strongly in the mid-IR. Here, dust spread out over the galaxy's length and heated by distributed star formation may remain at modest temperatures and reradiate more diffusely, helping to reconcile strong apparent reddening with weak hot-dust signatures. Limited MIRI detections for many individual sources, together with stacking constraints, suggest that any hot-dust component is modest compared to classical unobscured quasars in a large fraction of the population. In our geometry, an extended and clumpy distribution of dust and gas can reduce the prominence of a compact, intensely heated mid-IR peak, while still producing strong line-of-sight reddening and continuum suppression. Alternatively, as some authors have proposed, the obscuration may be dominated by very dense gas including a substantial hydrogen n=2 population (e.g., Inayoshi & Maiolino 2025), producing Balmer-continuum absorption without necessarily generating strong infrared emission. A sightline down the prolate galaxy naturally provides the high gas column density required in such scenarios, analogous to looking through a dense, gas-rich filament. In either case (dust or gas), the prolate geometry delivers a high column along the line of sight and a high projected surface density without requiring an unrealistically high volume density, and it remains consistent with current ALMA limits on dust-reprocessed luminosity (Casey et al. 2025).

In summary, the end-on view provides a natural route to the V-shaped SED: (i) an extended, end-on column strongly reshapes the continuum, making the rest-optical extremely red and depressing the UV-to-optical ratio, while (ii) UV light from less-obscured near-side star-forming regions can remain directly visible, preserving a bright, blue UV component. The resulting combination can mimic a strong break-like feature around the Balmer/4000 Å region and produce the observed V-shaped curvature without invoking a finely tuned mixture of unrelated components. Instead, geometry plus a clumpy, normal ISM in a prolate system can yield the observed SED shape in a physically plausible way.

**4.4 Large Apparent Balmer Line Widths: Orientation-Enhanced Kinematics and Columns plus Electron Scattering**

LRDs frequently exhibit very broad Balmer emission lines (e.g., $H_\alpha$ or $H_\beta$ with FWHM ~ 1000–3000 km s$^{-1}$, e.g. Kokorev et al. 2024; Kocevski et al. 2024; Rusakov et al. 2026), far exceeding the ~100 km s$^{-1}$ linewidths typical of star-forming galaxies at similar redshifts. Such widths are often interpreted as arising from a classical AGN broad-line region (BLR). However, for the low stellar masses inferred for many LRD hosts, a straightforward BLR virial interpretation can imply black holes that are over-massive relative to their galaxies. Our end-on prolate scenario suggests that the observed line widths need not be a single-component virial signature. Instead, two effects can act in synergy: (i) orientation-enhanced galaxy-scale kinematics (setting a broad base at the few × 100 km s$^{-1}$ level), and (ii) radiative-transfer broadening via electron scattering in a dense ionized medium (producing the extreme wings and pushing the apparent FWHM to > 1000 km s$^{-1}$).

(1) Geometric amplification of anisotropic kinematics (broad base).
If high-redshift low-mass galaxies are intrinsically prolate and dynamically warm/hot, their velocity field may be anisotropic, with a velocity-dispersion tensor elongated along the major axis ($\sigma_{\parallel} > \sigma_{\perp}$) and/or with significant streaming motions (inflows/outflows) preferentially aligned with the elongated structure. In that case, the line-of-sight dispersion depends on the viewing angle θ between the major axis and the line of sight, approximately as: $\sigma_{los}^2(\theta) \approx \sigma_{\parallel}^2 \cos^2(\theta) + \sigma_{\perp}^2 \sin^2(\theta)$.
For an end-on view (θ ~ 0°), $\sigma_{los} \sim \sigma_{\parallel}$, whereas for a side-on view (θ ~ 90°), $\sigma_{los} \sim \sigma_{\perp}$. This provides a natural geometric amplification of the observed line width by a factor of order $\sigma_{\parallel}/\sigma_{\perp}$, without requiring an unusually massive or exotic object. In addition, the end-on projection superposes multiple emitting regions distributed along the elongated galaxy, so that velocity gradients and multiple kinematic components along the line of sight can blend into a single broadened profile. Together, these effects are expected to generate line widths at the few × 100 km s$^{-1}$ level, plausibly explaining the "broad base" component in some LRD spectra.

It should be noted that this mechanism requires anisotropic random motions and/or streaming aligned with the major axis. If the dominant ordered motion were rotation about the major axis, an end-on view would contribute little Doppler broadening from that ordered component. Therefore, the observed broad bases favor a significant contribution from anisotropic dispersion and/or streaming along the elongated structure.

(2) Electron scattering in a dense ionized cocoon (extreme wings and apparent FWHM).
To explain the most extreme widths (FWHM > 1000 km s$^{-1}$) without requiring an over-massive black hole, we invoke electron (Thomson) scattering in a dense ionized medium surrounding the nuclear line-emitting region, following the interpretation advanced by Rusakov et al. (2026). In this picture, the observed broad Balmer profiles are dominated by prominent, approximately symmetric scattering wings superposed on a much narrower intrinsic line core, implying that the measured FWHM is not a direct virial tracer of the black-hole potential. Electron scattering with an effective Thomson optical depth of order unity (i.e., requiring Compton-thick electron columns, $N_e \sim 10^{24}$ cm$^{-2}$) can naturally broaden line photons to the observed thousand-km/s scale, thereby decoupling the apparent line widths from the black-hole virial velocity and alleviating the tension with low host stellar masses (Rusakov et al. 2026).

Crucially, the end-on prolate geometry provides a natural path to large column density. Because the line of sight in an end-on view traverses the maximum path length through the galaxy's central regions, both the dust/gas absorbing column and the ionized scattering column are maximized. Thus, the same geometric configuration that produces strong continuum reddening and suppression can also preferentially produce the large Thomson optical depth needed for electron-scattering broadening. This also offers a straightforward explanation for the unusually large Balmer equivalent widths in many LRDs: if the continuum is heavily attenuated while a fraction of line photons are scattered into the line of sight, the line-to-continuum contrast can be strongly enhanced.

In short, the broad lines do not necessarily require an over-massive BH; they can result from a combination of (a) anisotropic stellar motions (providing a broad base to the velocity distribution) and (b) scattering in a dense environment (fattening the line wings). The end-on prolate geometry is conducive to both: it maximizes the line-of-sight component of any anisotropic motion and ensures a long path through dense gas for scattering. Our model basically provides the physical setting for such a scenario to occur naturally, without requiring each LRD to be a special, rare case of an extremely massive black hole. Instead, a normal seed black hole in a dense, elongated early galaxy could appear as a "mini-quasar" when viewed end-on.

**4.5 High luminosity and energy output**

One might worry that interpreting LRDs as star-forming galaxies (rather than quasars) would demand implausibly high star formation rates to account for their brightness. Not necessarily, and the end-on prolate scenario can actually ease this concern.

Because LRDs are red, their observed emission is dominated by rest-frame optical/near-IR light. As a reference point, an object with F444W ~ 26 (AB) at z ~ 6 corresponds to rest-frame wavelength ~0.63 micron and a monochromatic luminosity $\nu L_\nu$ of order a few × $10^{43}$ erg/s. This is luminous, but it is not automatically "quasar-level" when interpreted as stellar emission plus modest reprocessing. A more direct way to assess the required power is to use the UV- and IR-based star formation calibrations rather than treating a single-band $\nu L_\nu$ as a bolometric luminosity. Photometrically selected LRDs span observed UV absolute magnitudes roughly $M_{UV}$ ~ −17 to −22 (Kokorev et al. 2024). Taken at face value without extinction correction, these correspond to $SFR_{UV}$ of about 0.3–30 $M_\odot$/yr (the exact values depend mildly on the adopted IMF and calibration). Independent constraints from ALMA 1.3 mm observations provide a strong upper bound on dust re-radiation: stacking 60 LRDs yields a non-detection and implies an upper limit on the total infrared luminosity less than or about $10^{11}$ $L_\odot$ (Casey et al. 2025), corresponding to $SFR_{IR}$ less than or about 10–20 $M_\odot$/yr for standard IR-to-SFR conversions. Combining these two pieces, the total star formation rate implied by UV plus the maximal allowed IR is typically at the level of a few to a few tens of $M_\odot$/yr, and even for the UV-bright end it is more naturally tens rather than hundreds to thousands. These numbers are well within the normal range for star-forming main-sequence galaxies of $M_* \sim 10^{9-10}$ $M_\odot$ at high redshift. In other words, from a purely energetic standpoint, LRDs do not necessarily harbor "overpowered" energy sources, their bolometric outputs can be accounted for by vigorous but not extreme star formation, especially given the tight ALMA limits on reprocessed dust emission.

Thus, the end-on prolate model is energetically feasible and provides a natural explanation for why LRDs can look "AGN-like" without requiring an intrinsically extreme bolometric power source. In our picture, the unusually high surface brightness and compact appearance are primarily projection effects: when a prolate galaxy is viewed close to end-on, a substantial fraction of its stellar light (and any nebular emission) is concentrated into a much smaller projected area. At the same time, the same end-on orientation maximizes the line-of-sight column through gas and dust (or dense neutral gas), which can strongly suppress the observed UV/blue continuum and reshape the SED. This combination, enhanced projected surface brightness plus continuum suppression, naturally increases the prominence of emission lines relative to the continuum and can make the object resemble an AGN in broadband colors and spectroscopy, even if the underlying total luminosity is comparable to that of normal star-forming main-sequence galaxies at the same epoch.

Large-area samples reinforce this point. For example, COSMOS-Web reports hundreds of LRD candidates and emphasizes that treating their continuum emission as AGN-dominated and applying standard quasar templates and bolometric corrections can lead to very large inferred $L_{\rm bol}$ values, whereas multi-wavelength constraints and alternative decompositions allow much lower bolometric outputs consistent with vigorous but not extreme star formation (Akins et al. 2025). This population-scale perspective strengthens the conclusion that the energy budget alone does not require quasar-level hidden power for the bulk of the LRD population.

This also clarifies why some AGN-based analyses inferred very high bolometric luminosities for LRDs. Those large bolometric luminosity values are not directly demanded by the observed UV+IR energy budget; rather, they can arise when one assumes an AGN-dominated SED and then applies large de-reddening corrections and standard quasar bolometric corrections, or converts broad-line luminosities to continuum using normal-AGN equivalent widths. If the observed optical continuum contains a significant host-galaxy contribution and if strong emission lines contaminate broad photometric bands, the AGN-based bolometric inference can be further biased high. Consistent with these concerns, Greene et al. (2026) use multi-wavelength constraints to derive empirically calibrated bolometric corrections for LRDs and argue that published bolometric luminosities can be reduced by up to about an order of magnitude. Once these assumptions are relaxed and empirical constraints (including the ALMA limits) are enforced, the implied total power of many LRDs is consistent with vigorous but not extreme star formation, optionally with a modest AGN contribution in some cases.

**4.6 Summary**
In summary, the End-on Prolate Galaxy model simultaneously addresses: the rarity (only a small orientation subset of an otherwise common population), the extreme compactness and surface brightness (a projection effect), the red color and a V-shaped SED (extinction through a long path), the lack of strong IR dust emission (geometric distribution of dust/gas or possibly HI absorption instead of dust), the broad line profiles (anisotropic orbits seen end-on plus electron scattering in a dense medium along the line of sight), the overall luminosity and energy output (consistent with vigorous but not extreme star formation, with a modest AGN contribution in some cases). All these are achieved within a physically plausible scenario for early galaxies, dovetailing with independent evidence that early low-mass galaxies are elongated and dynamically hot.

**5. Implications, Predictions, and Future Tests**
If the prolate unified model is correct, it has significant implications for our understanding of early galaxy evolution and offers several falsifiable predictions.

### 5.1 Connection between LRDs and normal galaxies

In this model, LRDs are not a separate class of objects powered by fundamentally different physics. Instead, they represent an orientation-selected subset of otherwise normal high-redshift galaxies, preferentially those with intrinsically prolate shapes and large line-of-sight columns when viewed close to the major axis. Their evolution should therefore be continuous with that of the broader low-mass galaxy population at the same epoch.

The model thus predicts an evolutionary trend: the LRD phenomenon should become less common toward lower redshift as the typical intrinsic shapes of low-mass galaxies evolve from elongated/prolate toward more oblate, rotationally supported configurations. This expectation is consistent with the broader empirical and theoretical picture that low-mass galaxies at low redshift are predominantly oblate and disk-like, whereas high-redshift low-mass systems show a much larger fraction of elongated/prolate morphologies (e.g., Pandya et al. 2024; Wang et al. 2024). Under this interpretation, the rarity of obvious LRD analogs in the local universe is a natural consequence of shape evolution plus the fact that the special end-on configuration becomes less frequent as the global prolate fraction declines.

### 5.2 Kinematics tests

A critical test of the model will come from spatially resolved kinematics, for example using JWST/NIRSpec IFU and, in the future, ELT-class integral-field spectroscopy. The key point is that the sign of the inclination dependence of the observed line-of-sight velocity width differs between prolate, dispersion-dominated systems and oblate, rotation-dominated disks.

(1) Prolate, dynamically warm/hot case (our preferred regime at high redshift). If low-mass galaxies are intrinsically prolate with anisotropic motions and/or streaming preferentially aligned with the major axis, then the observed line-of-sight dispersion should be largest when the system is viewed close to end-on (i.e., the major axis aligned with the line of sight, which corresponds to the LRD configuration). In that case, the measured linewidth is expected to trace the larger component of the velocity ellipsoid ($\sigma_{\parallel}$). When the same galaxy is viewed more side-on (appearing elongated/flattened on the sky), the line-of-sight dispersion should be smaller on average, closer to $\sigma_{\perp}$. This yields a clear, testable correlation: at fixed mass, the roundest and most compact-looking objects (end-on projections) should preferentially show the largest linewidths and strongest continuum suppression.

(2) Oblate, dynamically cold disk case (a contrasting hypothesis). If instead the relevant galaxies are intrinsically oblate and rotation-dominated, then "end-on" corresponds to a face-on view along the angular-momentum axis, and the projected rotation along the line of sight is minimized. In that case, the observed linewidth should be smallest for face-on systems and largest for edge-on systems, where projected rotation and/or disk thickness contribute most strongly to the line-of-sight velocities. Therefore, the oblate-disk picture predicts the opposite inclination trend compared to the end-on prolate picture.

Future tests are straightforward: a consistent "prolate signature" would be a positive correlation between linewidth and projected roundness/compactness, whereas a "disk signature" would be a negative correlation driven by inclination-dependent projected rotation.

**5.3 X-ray and variability**

A key advantage of the end-on prolate interpretation is that it does not require LRDs to be powered predominantly by an AGN. If a substantial fraction of LRDs are star-formation dominated systems whose AGN-like appearance is largely driven by projection (high apparent surface brightness) and column effects (continuum suppression and extreme line-to-continuum contrast), then weak X-ray emission and weak short-timescale variability are natural expectations. In this case, the rest-UV/optical continuum is primarily stellar (and/or nebular) and should be comparatively stable, with any stochastic variability strongly diluted by the host-galaxy light.

These expectations are broadly consistent with current observational constraints. Most spectroscopically confirmed LRDs are not individually detected in X-rays in the deepest Chandra fields, and X-ray stacking analyses indicate that the population is X-ray weak compared to what would be expected from low-redshift type-I AGN scaling relations at the same optical or line luminosities (e.g. Yue et al. 2024).

Time-domain tests provide an additional discriminator. Multi-epoch JWST imaging of large LRD samples shows that, as a population, LRDs do not exhibit strong variability in the rest-frame UV-optical on the currently accessible baselines, although a minority of sources do show significant variability at the level of a few tenths of a magnitude (e.g. Zhang et al. 2025). Targeted analyses of individual broad-line LRDs similarly report stringent non-detections of variability in several cases, which may pose tension for a simple unobscured, continuum-dominated type-1 AGN interpretation unless one invokes a host-dominated continuum, scattering-dominated emission, or intrinsically suppressed variability.

If a buried AGN is present in some LRDs, the end-on prolate geometry provides a natural reason why both X-ray and variability signatures can be muted. The same line of sight that maximizes the optical/UV attenuation can also correspond to very large gas columns that suppress the direct X-ray emission, while the observed UV-optical continuum can be dominated by host-galaxy light or reprocessed/scattered components that dilute or smooth intrinsic AGN variability. In this hybrid picture, one expects that strong variability should be confined to a small subset of LRDs where the AGN contributes substantially to the observed continuum, consistent with the current finding that only a minority of LRD candidates are strongly variable.

Future tests are straightforward: (i) deeper X-ray observations and larger stacking samples should tighten constraints on the typical X-ray-to-optical ratio of LRDs and its dependence on color/compactness, and (ii) expanded multi-epoch JWST monitoring should determine whether variability correlates with the degree of continuum suppression, line equivalent width, or apparent compactness, as expected if orientation-driven columns control the visibility of any nuclear component.

## 5.4 Environment and cosmic-web orientation effects

In the end-on prolate unified picture, LRDs are primarily an orientation-selected subset of intrinsically elongated (prolate) high-redshift low-mass galaxies. In its minimal form, this hypothesis does not require any special environmental trigger: at fixed stellar mass (or rest-frame optical luminosity), LRDs should inhabit similar dark matter haloes and have similar large scale bias as the parent galaxy population. Therefore, a key prediction is that on Mpc scales the clustering of LRDs should be broadly consistent with that of mass-matched normal galaxies, rather than resembling the strong bias expected for a population dominated by luminous quasars.

Current observational indications of LRD environments are not yet fully convergent and appear sensitive to sample definition, redshift uncertainties, and the scale used to quantify environment. Some analyses report that LRDs tend to lie in relatively sparse projected environments and show weak clustering compared to typical AGN expectations (e.g. Carranza-Escudero et al. 2025), which is broadly compatible with an orientation-based interpretation. On the other hand, individual spectroscopically confirmed systems demonstrate that at least some LRDs can reside in pronounced galaxy over-densities (e.g. Schindler et al. 2025), and there are emerging hints for an excess of very close (kpc-scale) LRD-LRD pairs (e.g. Tanaka et al. 2024). At present, these results should be regarded as complementary rather than contradictory: available samples remain limited, cosmic variance can be severe at $z > 5$, and projected-density estimators can be affected by selection and blending in crowded regions.

Within our model, the most natural way for large scale structure to modulate the observed LRD incidence is through orientation statistics: if prolate galaxies preferentially align their major axes with local filaments, then the fraction of galaxies observed close to end-on could vary from field to field depending on the filament orientation relative to the line of sight. While directly reconstructing the cosmic web and measuring such orientation effects at $z > 3$ is challenging with current data, this possibility provides a plausible route by which "environment" could enter an otherwise geometry-driven phenomenon, primarily by enhancing field-to-field variance and small-scale apparent associations.

Finally, we note a separate possibility that could contribute to apparent environmental diversity: in very dense environments, dissipative processes (e.g., gas-rich compaction) may form intrinsically compact, three-dimensional spheroids that could appear as small, red, high-surface-brightness objects. Whether such systems satisfy the current LRD selection criteria, and how strongly they contaminate LRD samples, remains unclear; if present, they would likely differ from the end-on prolate subset in detailed morphology (e.g., the presence or absence of faint elongated outskirts), kinematics, and multi-wavelength energy constraints.

## 6. Conclusion

We have proposed an end-on prolate interpretation for JWST Little Red Dots (LRDs) and argued that it can also place the unexpectedly abundant population of high-redshift "disk-like" galaxies into a single geometric framework. The central idea is simple: if many low-mass galaxies at $z > 3$ are intrinsically prolate, then viewing angle alone naturally produces both flattened, elongated disk-like appearances (side-on and intermediate orientations) and a small subset of extremely compact, high-surface-brightness objects (near end-on), i.e., LRDs. This picture does not invoke new fundamental physics; it instead emphasizes geometry and orientation as key ingredients for interpreting galaxies in the young universe.

This interpretation is motivated by converging theoretical and observational evidence that high-redshift low-mass galaxies are often elongated and dynamically warm/hot rather than classic rotation-supported oblate disks. Simulations have long suggested that early low-mass systems can be dispersion-dominated and prolate. Observationally, two largely independent approaches point in the same direction: three-dimensional structural modeling of individual galaxies indicates a substantial prolate fraction (Pandya et al. 2024), and statistical analyses of axis-ratio distributions at $3 < z < 7$ favor prolate intrinsic shapes while disfavoring a predominantly oblate disk population (Wang et al. 2024). A key concern is whether face-on disks might be systematically missed by surface-brightness selection; however, completeness tests in deep JWST imaging indicate that, over the relevant magnitude range, the data are close to complete even for large, face-on disks, making it unlikely that a dominant hidden face-on disk population is driving the inferred prolate preference (Pandya et al. 2024). Taken together, theory and current observations support a scenario in which prolate, dynamically warm/hot low-mass galaxies are common at $z > 3$. If so, a common prolate parent population *inevitably* yields many apparent "disks" and a small end-on tail that appears as LRDs.

Within this framework, the end-on prolate model can simultaneously account for multiple defining LRD properties with minimal additional assumptions: their rarity (a small orientation-selected subset), extreme compactness and high surface brightness (projection), red colors and a V-shaped SED (enhanced line-of-sight columns), the limited dust-reprocessed IR power (geometry and/or gas absorption), broad line profiles (orientation-enhanced kinematics together with electron scattering along the line of sight), and the overall luminosity and energy budget (consistent with vigorous but not extreme star formation, with a modest AGN contribution in some cases). In this view, LRDs need not represent a new exotic population; they are a natural manifestation of the same high-redshift galaxy population revealed by JWST under a special viewing geometry.

The model yields several near-term, falsifiable predictions, especially from kinematics constraints, which can validate or rule out this geometric interpretation. If validated, it would imply that the early universe was dominated by elongated, dynamically warm/hot galaxies, a picture that differs markedly from the classic expectation of early, rapidly settling disks.

**Acknowledgements**
Y.P. acknowledges support from the National Natural Science Foundation of China (NSFC) under grant Nos. 12125301, 12192220, and 12192222, and from the New Cornerstone Science Foundation through the XPLORER PRIZE.

**References**
Akins, H. B., et al. 2025, ApJ, 991, 37
Allgood, B., Flores, R. A., Primack, J. R., et al. 2006, MNRAS, 367, 1781
Baggen, J. F. G., et al. 2025, arXiv:2512.03239

Carranza-Escudero, M., et al. 2025, ApJL, 989, L50
Casey, C. M., et al. 2025, ApJL, 990, L61
Ceverino, D., Primack, J. & Dekel, A., 2015, MNRAS, 453, 408
Ferreira, L., et al. 2022, ApJL, 938, L2
Greene, J. E., et al. 2026, ApJ, 996, 129
Inayoshi, K., & Ho, L. C. 2025, arXiv:2512.03130
Inayoshi, K., & Maiolino, R. 2025, ApJL, 980, L27
Lapiner, S., Dekel, A., Freundlich, J., et al. 2023, MNRAS, 522, 4515L
Kocevski, D. D., et al. 2025, ApJ, 986, 126
Kokorev, V., et al. 2024, ApJ, 968, 38
Mo, H. J., Mao, S., & White, S. D. M. 1998, MNRAS, 295, 319
Pandya, V., et al. 2024, ApJ, 963, 54
Pandya, V., Loeb, A., McGrath, E. J., et al. 2025, ApJ, 986, 72
Pozo, A., Broadhurst, T., Emami, R., et al. 2026, Nat. Astron., 10, 306
Rusakov, V., et al. 2026, Nature, 649, 574
Schindler, JT., Hennawi, J.F., Davies, F.B. et al. 2025, Nat. Astron., 9, 1732
Tanaka, T. S., Silverman, J. D., Shimasaku, K., et al. 2024, arXiv:2412.14246
Tomassetti, M., et al. 2016, MNRAS, 458, 4477
Wang, B., Peng, Y., Cappellari, M., et al. 2024, ApJL, 973L, 29
Yue, M., et al. 2024, ApJL, 974L, 26
Zhang, Z., et al. 2025, ApJ, 985, 119